\documentclass[letterpaper, 10 pt, conference]{sty/ieeeconf}
\pdfoutput=1

\IEEEoverridecommandlockouts

\usepackage{graphicx}
\usepackage{amsfonts} 
\usepackage{enumerate} 
\usepackage{sty/my_SIunits}
\usepackage{sty/my_acronyms}
\usepackage{amsmath}
\usepackage{bm}
\usepackage{dblfloatfix}

\acrodefplural{DOF}{degrees-of-freedom}
\usepackage{cite}	
\usepackage{amssymb}    
\usepackage[usenames]{color}
\usepackage{multirow} 
\usepackage{balance} 

\graphicspath{{figs/}}
\newcommand{\ignore}[1]{}

\usepackage{stackengine}
\stackMath

\usepackage[backref]{hyperref}
\makeatletter
\newcommand*\titleheader[1]{\gdef\@titleheader{#1}}  
\AtBeginDocument{%
  \let\st@red@title\@title
  \def\@title{%
    \vskip-3.5em\bgroup\normalfont\large\centering\@titleheader\par\egroup
    \vskip2.0em\st@red@title}
}
   \let\NAT@parse\undefined
\makeatother

\title{Uniform Complete Observability of Mass and Rotational Inertial Parameters in Adaptive Identification of Rigid-Body Plant Dynamics\\        
}
\titleheader{2021 IEEE International Conference on Robotics and Automation (ICRA 2021) \\
Xi'an, China, May 30 - June 5, 2021}

\author{Tyler M. Paine$^1$, and Louis L. Whitcomb$^1$
  \thanks{We gratefully acknowledge the support of the DoD Science, Mathematics, And Research For Transformation (SMART) Defense Scholarship Program,  and   
    the National Science Foundation under award IIS-1909182.}
   \thanks{1. Department of Mechanical
     Engineering, Johns Hopkins University, Baltimore, MD 21218, USA 
     {\tt\small tpaine1,llw@jhu.edu} } }

\begin{document}

\maketitle


\begin{abstract}

This paper addresses the long-standing open problem of observability
of mass and inertia  plant parameters in the \ac{AID} of second-order
nonlinear models of 6 degree-of-freedom
rigid-body dynamical systems subject to externally applied forces and moments.  
Although stable methods for \ac{AID} of plant parameters
for this class of systems, as well numerous approaches to stable
model-based direct adaptive trajectory-tracking control of such
systems, have been reported, these studies have been unable to prove
analytically that the adaptive parameter estimates converge to the
true plant parameter values.
This paper reports necessary and sufficient conditions for the uniform
complete observability (UCO) of 6-DOF plant inertial parameters for a
stable adaptive identifier for this class of systems.  When the UCO
condition is satisfied, the adaptive parameter estimates are shown to
converge to the true plant parameter values.  
To the best of our knowledge this is the first reported proof for this class of systems
of UCO of plant parameters and for convergence of adaptive parameter
estimates to true parameter values.

We also report a numerical simulation study of this \ac{AID}
approach which shows that (a) the UCO condition can be met for
fully-actuated plants as well as underactuated plants with the proper
choice of control input and (b) convergence of adaptive parameter
estimates to the true parameter values.  We conjecture that this
approach can be extended to include other parameters that appear rigid
body plant models including parameters for drag, buoyancy, added mass, bias, and
actuators.

\end{abstract}

\section{Introduction} \label{sec.intro}

\newcommand{\bea}{\begin{eqnarray}}
\newcommand{\eea}{\end{eqnarray}}

\newcommand{\be}{\begin{equation}}
\newcommand{\ee}{\end{equation}}

\newcommand{\R}{{\rm I\!R}}

This paper addresses the long-standing open problem of observability 
of plant parameters 
and convergence of plant parameter estimates in the \acf{AID} of second-order rigid-body dynamical systems.
We address the problem of
\ac{AID} of mass and rotational inertia parameters for 6
degree-of-freedom (6-DOF) rigid-body plants subject to externally
applied forces and moments.  
Examples of such plants include 
aerospace, marine, and terrestrial robotic vehicles, which operate in SE(3). 
The plant models developed for these robotic systems could certainly include additional terms to account for gravitational forces, hydrodynamic drag, etc. 
However, these SE(3) plant models all share the nonlinear Coriolis terms, which include the inertial parameters and have the effect of coupling motion between different degrees of freedom.  
For this reason, it is important to first develop the \ac{AID} and observability conditions for parameters which are common among many members of this class of dynamical systems: the mass and rotational inertia parameters.  
We conjecture that this approach reported herein can be extended to include other parameters that appear rigid-body plant models, including parameters for drag, buoyancy, bias, added mass, and actuators, and to other classes of mechanical systems.

Numerous approaches have been reported in the literature for stable
\ac{AMBTTC} this class of systems, as well as a (very) few approaches
for \ac{AID} of plant parameters for this class of systems.  Although
these studies have shown stability of the adaptive parameter estimates,
none has able
to prove analytically
that the plant parameters are observable, or that
the adaptive parameter estimates converge to the
true plant parameter values.
Many of these studies, however, report
numerical simulations and/or experimental evaluations
which demonstrated the convergence of the parameter estimates to
the true parameter values --- strongly suggesting that the
parameters are, in fact, observable.

{\it Relation of \acf{AID} to \acf{AMBTTC}:}
The present study \ac{AID}
does not address the problem of
  simultaneously performing adaptive model identification {\it and}
  model-base trajectory tracking control, i.e.
\acf{AMBTTC}. 
There are numerous previously reported results for
\ac{AMBTTC}
of mechanical systems, in which the plant
state is controlled to follow a {\it a priori} known smooth trajectory.
\ac{AMBTTC}, however, will not work when either
$(i)$ the plant is underactuated or
$(ii)$ the plant is controlled by an existing controller that cannot be modified,
as is often the case with  commercially available robot systems, or
$(iii)$ the plant is under open-loop control, as is often the case for system
identification experiments in the early stages of vehicle development.

Unlike \ac{LS} approaches to identification of general rigid-body plant parameters, which require actuator force/moment, position, velocity, and acceleration signals, previously reported \ac{AID} methods require only actuator force/moment, position, and velocity signals - but do not require acceleration signals. The fact that these \ac{AID} methods do not require acceleration signals is a significant advantage, since in practice acceleration signals are often numerically calculated from velocity data - a process that is extremely sensitive to noise.  For the class of plants addressed herein, without any buoyancy or other gravitational forces, the \ac{AID} approach requires only the signals of velocity and applied force/moment.

The remainder of this paper is organized as follows:
Section \ref{sec.litReview} provides a review and discussion of previous related work.
Section \ref{sec.aid} reviews stable \ac{AID} of mass and inertial parameters in rigid-body plants.
Section \ref{sec.uco_def} then reports the definition of Uniform Complete Observability (UCO) and the error dynamics system for rigid-body motion.
Section \ref{sec.impl} reports a  practical numerical implementation and methods to identify the observable parameter subspace.
Section \ref{sec.simStudy} reports the methodology used in the numerical simulation study and the simulation results.
Section \ref{sec.conc} concludes.

\section{Literature Review} \label{sec.litReview}
Most previously reported plant parameter identification methods 
employ one of two general approaches: $(i)$ least-square
linear regression of experimental data or $(ii)$ adaptive model-based
trajectory-tracking control.
A variety of previously reported studies have employed least-squares,
total least-squares, or extended Kalman filters to identify plant
parameters entering linearly into the plant equations of motion for
robot manipulators
\cite{khosla&kanade,Atkeson1986,an&atkeson&hollerbach,armstrong&khatib&burdick,swevers.2007}, 
\acp{UUV} \cite{caccia.joe2000,alessandri.ca98,martinROV_ID_OE}, and
spacecraft \cite{Norman2011,Keim2006}.  Khalil and Dombre provide a
detailed overview \cite{Khalil2002}.

The concept of persistent excitation and its relationship to the convergence of parameter error is well known and widely utilized in the field of adaptive systems
\cite{narendra_adaptive_book_1989,sastryadaptive}.
Recent discoveries in geometric characterizations of persistently exciting signals in neutrally stable autonomous systems were reported in \cite{padoan2017geometric}.  Padoan \textit{et al.} report a geometric characterization of persistent excitation condition for signals generated by time-invariant, continuous-time, dynamical systems without a forcing term. Regardless, the algebra required to compute the rank condition presented by Padoan \textit{et al.} is problematic for 6-DOF rigid body systems, and the rank condition is only a sufficient condition for persistent excitation.  Furthermore, there was no consideration of a forcing term in the system definition.  

Klaus R\"{o}benack and Pranay Goel in \cite{robenack2009IJSS} report a combined observer and filter-based approach to identify unknown parameters of non-linear systems which does not require persistent excitation.  However, this approach is limited to a particular class of systems that are affine in the unknown parameters, and the rigid-body plant model described by the Newton-Euler equations cannot be transformed to match.

Besan\c{c}on reported a study of a class of nonlinear plants and parameter observers,
and established conditions for \ac{UCO} of the plant parameter estimates \cite{besanccon2000remarks}.
Jenkins et al. reported the relation of \ac{PE} and \ac{UCO} for scalar adaptive plants, which  implies exponential stability in the large (ESL) \cite{Jenkins2018SIAM}.
In these approaches, the dynamical system of the error coordinates is needed and, in the case of rigid body dynamics, this system will be time-varying, and thus the observability condition is \emph{uniform} for a given sliding time window $[t_0, t_0 + \delta] \ \ t_0, \delta \in {\rm I\!R}_+$.  A practical application of this technique was employed by Spielvogel and Whitcomb in \cite{spielvogel2020IJRR} to show asymptotic convergence of the sensor bias and observer to the true values.
Spielvogel and Whitcomb show the equivalence  between the signals having the property of persistent excitation and UCO of the error system
(\hspace{-0.6mm}\cite{spielvogel2020IJRR} Lemma 1, proof in Appendix 1).  This equivalence is shown in \cite{Jenkins2018SIAM} for scalar adaptive plants.  This approach is used in the present paper to develop necessary and sufficient conditions for \ac{UCO} for \ac{AID} of  mass and inertial parameters of second-order nonlinear rigid-body dynamical systems.

\section{Stable Adaptive Identification of Mass and Inertial Parameters Rigid-Body Plants}
\label{sec.aid}

We consider the problem of inertial parameter identification for a simple 6-\ac{DOF} rigid-body dynamical system, where the mass matrix $M \in {\rm I\!R}^{6 \times 6}$ is positive definite diagonal, with no drag or gravitational terms, and the control forces and moments $\tau(t) \in {\rm I\!R}^6$ are known time varying signals. 
The most commonly utilized finite-dimensional \ac{EOM} about the \ac{COM} in body-frame coordinates for this rigid-body dynamical system is \cite{fossen2011handbook,Paine2018}

\begin{equation}
  \tau(t)= M \dot{v}(t)  + C(M,v(t)) v(t),
  \label{simpleRBSystem}
\end{equation}
where
$v(t) \in {\rm I\!R}^6 = [\nu(t); \omega(t)]$ is the body relative velocity vector, where
$\nu(t) \in {\rm I\!R}^3$ is the translational body velocity vector, and
$\omega(T) \in {\rm I\!R}^3$ is the angular velocity vector.
In the sequel we will drop the explicit time dependence of the body velocity signals for clarity.
The mass matrix is
\be
M = diag(m),
\ee
where
$
  m = [m_{11}; m_{22}; m_{33}; m_{44}; m_{55}; m_{66}]^T \in {\rm I\!R}^6,
$ and the $m_{ii}$ are positive mass and moment of inertia parameters.

  The Coriolis matrix  $C(M,v(t))$  is defined as
\begin{align}
C(M,v) =& \begin{bmatrix}
0 & -J(M_{11} \nu) \\
-J(M_{11} \nu) & -J(M_{22} \omega)
\end{bmatrix}, 
\end{align} 
where $M_{11} = diag([m_{11}; m_{22}; m_{33}])$, $M_{22} = diag([m_{44}; m_{55}; m_{66}])$, and $J: {\rm I\!R}^{3} \rightarrow {\rm I\!R}^{3 \times 3}$ is the skew symmetric operator.

This simplified model is widely applicable to many 6-\ac{DOF} rigid-body dynamical plants.  We acknowledge that the mass matrix $M$ is not strictly representative of the classic rigid-body model, where the translational  inertia of the body in each of the three cardinal directions is equal.  Instead, we address the more general problem of inertial parameter identification in models of systems which can have different quantities of inertia in the different degrees of freedom.  A classic example of such a system is an underwater vehicle; these systems are often modeled to include added-mass components in $M$. Clearly, the classic rigid-body dynamical system is a sub-class of the larger class of systems considered herein.

The stable adaptive identifier reported in \cite{icra2013mcfarland,2013mcfarlandThesis,2019HarrisThesis} can be
adapted to identify only the mass matrix $M$ of the simplified plant
(\ref{simpleRBSystem}).  In the present study we seek to determine the
persistent excitation required for the adaptive parameter estimates
to converge to their true values.

The stable adaptive parameter identifier reported in \cite{icra2013mcfarland}, when applied to the plant \ref{simpleRBSystem}
takes the form of an  adaptive identifier 
and an adaptive parameter update law.
The adaptive identifier plant takes the form
\be
\dot{\hat{v}} = \hat{M}^{-1} (-C(\hat{M},v)v + \tau(t)) - \mathcal{A} \Delta v,
\label{v_update} 
\ee
where $\hat{v} \in \R^6$ is the adaptive identifer plant velocity,
and
\be
\hat{M} = diag(\hat{m}),\\
\ee
is the estimated mass matrix where
$
  \hat{m} = [\hat{m}_{11}; \hat{m}_{22}; \hat{m}_{33}; \hat{m}_{44}; \hat{m}_{55}; \hat{m}_{66}]^T \in {\rm I\!R}^6,
$ and the $\hat{m}_{ii}$ are time-varying adaptive estimates of the  mass and moment of true inertia parameters $m_{ii}$.  The adaptation gain $\mathcal{A} \in {\rm I\!R}^{6 \times 6}$ is a positive definite diagonal matrix of adaptation gains.  
The adaptive parameter update law takes the form
\be
\dot{\hat{m}} = \Gamma_1 (v^T diag (ad(v)^T \Delta v) + (\dot{\hat{v}}+\alpha \Delta v)^T diag( \Delta v) )^T
\label{m_update}
\ee
where 
\begin{align}
ad\Bigg(\begin{vmatrix}
\nu \\
\omega \\
\end{vmatrix}\Bigg) =& \begin{bmatrix}
J(\omega) & 0_{3\times3} \\
J(\nu) & J(\omega)
\end{bmatrix}.
\end{align} 
and $\Gamma_1 \in {\rm I\!R}^{6 \times 6}$ is positive definite diagonal matrix of adaptation gains.  The use of positive definite diagonal gain matrices was first reported in \cite{2019HarrisThesis}.
The error coordinates for this system are
\be
\Delta v = \hat{v} - v \in \R^6, \;\;\;\; \Delta m = \hat{m} - m \in \R^6
\ee
where
$\Delta M = \hat{M} - M = diag(\Delta m) = diag(\hat{m})-diag(m) $.

The stability properties of this \ac{AID} algorithm are useful in the sequel.  For this reason, a quick proof using Lyapunov's direct method is presented. The reader is referred to \cite{2013mcfarlandThesis, 2019HarrisThesis, Paine2018} for more details.  
From (\ref{simpleRBSystem}) and (\ref{v_update}) we can write the error system dynamics as 
\begin{align}
M \Delta\dot{v}  =& \, M (\dot{\hat{v}} - \dot{v}) \label{eq:vel_error_express} \\ 
	=& \mathcal{A} M \Delta v - \Delta M  (\dot{\hat{v}} + \alpha \Delta v) - ad(v) \Delta M v. 
	  \nonumber
\end{align}
Consider the following Lyapunov function candidate 
	\begin{align}
	V(t) =& \frac{1}{2} \Delta v^T M \Delta v + \frac{1}{2} \Delta m^T \Gamma_1^{-1}\Delta m. \label{eq:6DOF_V_def_decoupled}
	\end{align}
	The time derivative of (\ref{eq:6DOF_V_def_decoupled}) is
	\begin{align}
	\dot{V}(t) =& \frac{1}{2} \Big( \Delta\dot{ v} ^T M \Delta v + \Delta v^T M \Delta\dot{v} \Big) + \Delta \dot{m}^T \Gamma_1^{-1} \Delta m. 
	\end{align}
	Substituting in (\ref{eq:vel_error_express}) yields
	\begin{align}
	\dot{V}(t) =& - \Delta v^T \mathcal{A} M \Delta v - \frac{1}{2} \Big( v^T \Delta M \psi_1 + \psi_2^T \Delta M \Delta v \Big) \nonumber \\
	& - \frac{1}{2} \Big( \Delta v^T \Delta M \psi_2 + \psi_1^T \Delta M v \Big) + \Delta \dot{m}^T \Gamma_1^{-1} \Delta m, 
	\end{align}
	where $\psi_1 = ad(v)^T \Delta v$, and $\psi_2 = \dot{\hat{v}}+ \mathcal{A} \Delta v$.
	This is equivalent to 
	\begin{align}
	\dot{V}(t) =& - \Delta v^T \mathcal{A} M \Delta v  \nonumber\\
	& -\Big( v^T diag(\psi_1) + \psi_2^T  diag(\Delta v) \Big) \Delta m  \nonumber \\
	& +  \Delta \dot{m}^T \Gamma_1^{-1} \Delta m.
	\end{align}
	Substituting in the parameter update law (\ref{m_update}) yields
	\begin{align}
	\dot{V}(t) =& - \Delta v^T \mathcal{A} M \Delta v. \label{eq:V_dot}
	\end{align}

        While the Lyapunov function (\ref{eq:6DOF_V_def_decoupled}) is positive definite in both $\Delta v$ and $\Delta m$,
        (\ref{eq:V_dot}) is only negative definite in $\Delta v$. Proof of boundedness of all signals (including $\hat{M}^{-1}$) is given in \cite{icra2013mcfarland,2013mcfarlandThesis}.
          From (\ref{eq:6DOF_V_def_decoupled}), (\ref{eq:V_dot}), and the boundedness of all signals,  we can conclude that
            $ \lim_{t\rightarrow \infty} \Delta v = 0$ and that
            $ \lim_{t\rightarrow \infty} \Delta \dot{m} = 0$,
          but we cannot conclude  $ \lim_{t\rightarrow \infty} \Delta m = 0$ \cite{icra2013mcfarland,2013mcfarlandThesis}.

          Although McFarland and Whitcomb in \cite{icra2013mcfarland} were unable to prove analytically, experimental studies therein showed that
          the \ac{AID} parameter estimates converged repeatably to reasonable values. Moreover, numerical simulation studies reported in \cite{2013mcfarlandThesis} showed that \ac{AID} parameter estimates converged repeatably  to the true parameter values. 

          In this paper, we  seek fill this lacunae to show analytically 
          that these plant parameters are observable, and  to show that the
          \ac{AID} parameter estimates converge true parameter values.

\section{Uniform Complete Observability of Error Dynamic System} 
\label{sec.uco_def} 

In this section we will apply the concept \acf{UCO} to this \ac{AID} of rigid body dynamical plants to prove analytically the conditions where $\lim_{t \rightarrow \infty} \Delta m = 0$. 

There is an equivalence between \acf{PE} and \ac{UCO}.  The relationship has been shown analytically for scalar plant models \cite{Jenkins2018SIAM}, for a class of nonlinear plants and associated parameter observers \cite{besanccon2000remarks}, and experimentally verified in the problem arising in sensor bias and attitude estimation in gyrocompass systems \cite{spielvogel2020IJRR}. In the case of rigid body dynamic systems (\ref{simpleRBSystem}), the error system for $\Delta v$ (\ref{v_update}) and $\Delta m$ (\ref{m_update}) can be algebraically rearranged to be written as
\begin{align}
\begin{vmatrix}
\Delta \dot{m}(t) \\
\Delta\dot{v}(t)
\end{vmatrix} =&
A_e(t)
\begin{vmatrix}
\Delta m(t) \\
\Delta v(t) 
\end{vmatrix}, \label{error_dynam} \\
y(t) =& C_e  
\begin{vmatrix} 
\Delta m(t) \\
\Delta v(t) 
\end{vmatrix},   \nonumber
\end{align}
where $A_e(t) \in {\rm I\!R}^{12 \times 12}$ and $C_e \in {\rm I\!R}^{6 \times 12}$.

With the error system in this form, we recall the definitions of \acf{PE} and \acf{UCO}:

\textbf{Definition: Persistent Excitation (PE) \cite{narendra_adaptive_book_1989,sastryadaptive}:} A matrix function $\mathcal{W}:\mathbb{R}^+\rightarrow\mathbb{R}^{m\times n}$ is persistently exciting (PE) if there exist $T,\alpha_1,\alpha_2>0$ such that $\forall t\geq0$:
\begin{align}
	\alpha_1I_m\geq\int_{t}^{t+T}\mathcal{W}(\tau)\mathcal{W}^T(\tau)\,d\tau\geq\alpha_2I_m
\end{align}
where $I_m\in\mathbb{R}^{m\times m}$ is the identity matrix.

\textbf{Definition: Uniform Complete Observability (UCO) \cite{sastryadaptive}:} 
The system $[A_e(t), C]$ is called uniformly completely observable (UCO) if there exists strictly positive constants $\beta_1$, $\beta_2$, $\delta$ such that $\forall t_0 \geq 0$  $\beta_2I \geq N(t_0,\delta) \geq \beta_1I$. $N(t_0, \delta)$ is the observability Gramian,
\begin{align}
N(t_0, \delta) = \int_{t_0}^{t_0+\delta} \Phi^T(\tau, t_0) C^T C \Phi(\tau, t_0) d\tau, \label{obsv_gram}
\end{align} 
and $\Phi(\tau, t_0)$ is the state transition matrix for $A_e(t)$ \cite{rugh}.

Clearly a system $[A_e(t), C]$, from (\ref{error_dynam}),  is UCO if and only if the signal $  \Phi^T(\tau, t_0) C^T $ is PE.

To construct the  $A_e(t)$ appearing in (\ref{error_dynam})
we proceed as follows:  Beginning with $\Delta v$ and the adaptive identifier velocity law (\ref{v_update}), and noting the identity $C(M,v)v = ad(v)Mv$, we can write
\begin{align}
\Delta\dot{v} =& \dot{\hat{v}} - \dot{v}, \\
\Delta\dot{v} =& \hat{M}^{-1} \Big(-ad(v)\hat{M}v + \tau(t)\Big) - \mathcal{A} \Delta v \nonumber \\
&- M^{-1} \Big(-ad(v)Mv + \tau(t)\Big), \\
\Delta\dot{v} =& [M^{-1} M ] \hat{M}^{-1}  \Big(-ad(v)\hat{M}v + \tau(t)\Big) - \mathcal{A} \Delta v \nonumber \\
&- M^{-1} \Big(-ad(v)Mv + \tau(t)\Big).
\end{align}
 
With the identity  $M \hat{M}^{-1} = I - \Delta M \hat{M}^{-1}$, \cite{icra2013mcfarland}, the error dynamics can be rearranged as
\begin{align}
\Delta\dot{v} =& M^{-1} [I - \Delta M \hat{M}^{-1}]  \Big(-ad(v)\hat{M}v + \tau(t)\Big) - \mathcal{A} \Delta v \nonumber \\
&- M^{-1} \Big(-ad(v)Mv + \tau(t)\Big), \\
\Delta\dot{v} =& M^{-1}\Big(-ad(v)\hat{M}v + \tau(t)\Big) \nonumber \\
&- M^{-1} \Delta M \hat{M}^{-1} \Big(-ad(v)\hat{M}v + \tau(t)\Big) \nonumber \\
&- \mathcal{A} \Delta v  - M^{-1} \Big(-ad(v)Mv + \tau(t)\Big).
\end{align}
Since $M = diag(m)$, then $Mv = diag(m)v = diag(v)m$.  
\begin{align}
\Delta\dot{v} =& M^{-1}\Big(-ad(v)\hat{M}v + ad(v)Mv\Big)  \\
&- M^{-1} \Delta M \hat{M}^{-1} \Big(-ad(v)\hat{M}v + \tau(t)\Big)- \mathcal{A} \Delta v,  \nonumber \\
\Delta\dot{v} =& M^{-1}\Big(-ad(v) diag(v) \Big)\Delta m \nonumber \\
 & - M^{-1} \bigg( diag \Big( \hat{M}^{-1} \big(-ad(v)\hat{M}v + \tau(t) \big) \Big) \bigg) \Delta m \nonumber \\
&- \mathcal{A} \Delta v. 
\end{align}

The dynamics of the error coordinate $\Delta m$ and the associated parameter update law (\ref{m_update}) can be rearranged as
\begin{align}
\Delta\dot{m}  =& \dot{\hat{m}}-\dot{m}  = \dot{\hat{m}},\\
\Delta\dot{m}  =& \Gamma_1 \big( (ad(v)^T \Delta v)^T diag(v) \big)^T  \\
& + \Gamma_1 \bigg( \Big(\hat{M}^{-1} \big(-ad(v)\hat{M}v + \tau(t)\big)\Big)^T diag( \Delta v) \bigg)^T, \nonumber \\
\Delta\dot{m}  =& \Gamma_1 \big( \Delta v^T ad(v) diag(v) \big)^T  \\
& + \Gamma_1 \bigg( \Delta v^T diag \Big(\hat{M}^{-1} \big(-ad(v)\hat{M}v + \tau(t)\big) \Big) \bigg)^T, \nonumber\\
\Delta\dot{m}  =& \Gamma_1 \big( ad(v) diag(v) \big)^T \Delta v \\
& + \Gamma_1   diag \Big(\hat{M}^{-1} \big(-ad(v)\hat{M}v + \tau(t)\big) \Big)  \Delta v.  \nonumber
\end{align}

Finally, the error system dynamics can be written as
\begin{small}
\begin{align}
\begin{vmatrix}
\Delta\dot{m}(t) \\
\Delta\dot{v}(t)
\end{vmatrix} =&
\underbrace{\begin{bmatrix}
0 & \Gamma_1 [-S(t)^T + D(t)] \\
 M^{-1} [S(t) -D(t)] & -\mathcal{A}
\end{bmatrix}}_{A_{e}(t)}
\begin{vmatrix}
\Delta m(t) \\
\Delta v(t)
\end{vmatrix}  \label{error_dynam_system1}\\
y(t) =& \underbrace{\begin{bmatrix}
0_{6 \times 6} & I_{6 \times 6} 
\end{bmatrix}}_{C_e}
\begin{vmatrix}
\Delta m(t) \\
\Delta v(t) 
\end{vmatrix}, \label{error_dynam_system2}
\end{align}
\end{small}
where
\begin{align}
  D(t) =& diag \Big(\hat{M}^{-1} \big(-ad(v)\hat{M}v + \tau(t)\big) \Big),  \label{eq:DiagError} \\
S(t) =& -ad(v) \ diag(v). \label{eq:Skew}
\end{align}

Given the AID of rigid body dynamics as described in Section \ref{sec.aid}, all signals are bounded and $lim_{t \rightarrow \infty} \Delta v = 0$, which implies that $lim_{t \rightarrow \infty} || y(t) || = 0$ of the error system (\ref{error_dynam_system1}) and (\ref{error_dynam_system2}). If the pair $[A_e(t), C_e]$ is UCO then $lim_{t \rightarrow \infty} || \Delta m(t) || = 0$ as desired. (Proof is Lemma 1 in Appendix 1 of \cite{spielvogel2020IJRR}).  Thus, under this condition, parameter error will asymptotically converge to zero.

\section{Practical Numerical Implementation and Identification of Observable Subspace}
\label{sec.impl}

This section reports a  practical numerical implementation and methods to identify the observable parameter subspace.
The error system (\ref{error_dynam}) is a linear time varying (LTV) system.  The discrete approximation solution for this system is given by
\begin{align}
\begin{vmatrix}
\Delta m_{k+1} \\
\Delta v_{k+1}
\end{vmatrix} =&
\tilde{A}_e(k)
\begin{vmatrix}
\Delta m_{k} \\
\Delta v_{k}
\end{vmatrix} \\
y_{k+1} =& C_e
\begin{vmatrix}
\Delta m_{k+1} \\
\Delta v_{k+1} 
\end{vmatrix}.
\end{align}
where $\tilde{A}_e(k) = e^{A_e(t_{k})\Delta t}$, and $C_e$ is as defined before.

The state transition matrix over the interval $[k,k_0], k>k_0$ is given by 
\begin{align}
\Phi(k,k_0) = \tilde{A}_e(k-1)\tilde{A}_e(k-2) \hdots \tilde{A}_e(k_0). 
\end{align}
If $t_0 = k_0$, and $t = k*\Delta t$  this is the discrete approximation of $\Phi(t, t_0)$ with discrete timestep $\Delta t$.  \cite{edward2010LinSys}  As a result the integral in (\ref{obsv_gram}) can be approximated as:
\begin{align}
\int_{t_0}^{t_0+\delta} &\Phi^T(\tau, t_0) C_e^T C_e \Phi(\tau, t_0) d\tau \\
\approx& \sum_{i = 1}^{N=\delta/\Delta t} \tilde{A}_e^T (k_{i-1}) C_e^T C_e \tilde{A}_e (k_{i-1}). 
\end{align}
The corresponding Observability matrix $O(k)$, \cite{rugh}, for this discrete system is defined as 
\begin{align}
O(k) =& \begin{vmatrix}
O_0(k) \\
O_1(k) \\
\vdots \\
O_{N-1}(k)
\end{vmatrix} \label{eq:obsmatrix},
\end{align}
where the block rows $O_{i}(k)$ of $O(k)$ are given by
\begin{align}
O_0(k) =& C_e \\
O_{i}(k) =& O_{i-1}(k+1)\tilde{A}_e(k), \quad i = 1,2,\hdots, N-1
\end{align}

If $rank(O) = n_2 < 12$, then some subspace of the error system is not \ac{UCO}.  This is common in underactated rigid body dynamics.  For example, a 6-DOF rigid body robot moving in 3-DOF planar motion has rotational inertia in the out-of-plane degrees-of-freedom.  However, this out-of-plane inertia is not observable given the definition of planar motion, and this limit to observability is related to the lack persistent excitation in those out-of-plane degrees-of-freedom.  This example is intuitive, but we find that using the definition of \ac{UCO} of the error system provides a more rigorous mathematical approach to determining the observability of plant parameters.  The theoretical result reported in Section \ref{sec.uco_def} is certainly valid in the general case of fully-actuated rigid body motion.  It is also valid for the many classes of underactuated rigid body motion. 

The Kalman observable canonical form can be used to identify the subspace of the error system that is observable. Specifically, we can define a state transition matrix, $T\in {\rm I\!R}^{n \times n}$, where $\Delta \phi  = T \begin{vmatrix} \Delta m & \Delta v \end{vmatrix}^T$, such that the result is a state transformation to the Kalman observable canonical form.  Let $T = \begin{vmatrix} T_o & T_{uo} \end{vmatrix}^T$, where $T_o$ consists of $n_2$ linearly independent rows of $O$, and $T_{ou}$ are added rows to complete the basis and ensure $T$ is non-singular.  This state transformation yields the observable subspace as
\begin{align}
\begin{vmatrix}
\Delta \phi_{o} \\
\Delta \phi_{uo} 
\end{vmatrix} =& T \begin{vmatrix}
\Delta m \\
\Delta v 
\end{vmatrix},
\end{align}
where $\Delta \phi_{o}$ is the observable subspace, and $\Delta \phi_{uo}$ is the unobservable subspace.  In practice, given the definition of matrix $C_e$ we find the first $6$ rows of $T$ are the linearly independent basis vectors $\hat{e}_i, \ \ i = 7, 8, \hdots, 12$.  Thus $\Delta \phi_{o} \supset \Delta v$.  However, we note that some rigid-body robots are not instrumented with sensors to measure the body relative velocities in every degree-of-freedom. If this is the case, the matrix $C_e$ will be different, but the same results follow. 

The observability of remaining error states $\Delta m$ depends on the remaining $n_2-6$ rows of $T_o$. It is not obvious to the authors how to determine the closed form solution for these rows.  It was possible to numerically calculate the entries in these rows during simulation studies.  From this calculation, the rows of $T_{ou}$ can be clearly assembled, and thus the unobservable error states $\Delta m$ can be clearly identified. These studies are reported in the next section.  Regardless, result of the state transformation is the system in the classic Kalman canonical observability form
\begin{align}
\begin{vmatrix}
\Delta \phi_{o_{k+1}} \\
\Delta \phi_{uo_{k+1}}
\end{vmatrix} =& \begin{bmatrix}
\bar{A_e}_{o} 	& 0 \\
\bar{A_e}_{21}  & \bar{A_e}_{uo}
\end{bmatrix} \begin{vmatrix}
\Delta \phi_{o_{k}} \\
\Delta \phi_{uo_{k}}
\end{vmatrix} \\
y_{k+1} =& \begin{bmatrix}
\bar{C_e}_o & 0
\end{bmatrix} \begin{vmatrix}
\Delta \phi_{o_{k+1}} \\
\Delta \phi_{ou_{k+1}}
\end{vmatrix},
\end{align}
where the pair $[\bar{A_e}_{o},\bar{C_e}_o]$ is \ac{UCO}.

\section{Numerical Simulation Performance Analysis} 
\label{sec.simStudy}  
We evaluated the proposed approach to \acf{AID} of mass and inertia parameters of rigid-body plants, Section \ref{sec.aid}, and the associated \acf{UCO} analysis, Section \ref{sec.uco_def}, in numerical simulation studies for three classes of rigid body plants that differ only in the number and configuration of ther actuation, as follows:

\begin{itemize}

\item[$\bullet$] \textbf{Class 0} \textit{- Fully Actuated Plant}:
Control authority is available in all 6-DOF. 

\item[$\bullet$] \textbf{Class 1} \textit{- Underactuated Plant:} 
  Control authority is available in the longitudinal ($x_1$), roll ($x_4$), pitch ($x_5$), and yaw ($x_6$) DOF, and is 
  not available in the lateral ($x_2$) or vertical ($x_3$) DOF.  This is typical of many in torpedo shaped underwater vehicles and aerospace vehicles equipped with a single main propusor and control surface fins. 

\item[$\bullet$] \textbf{Class 2 } \textit{- Underactuated Plant:}
  Control authority is only available in the longitudinal ($x_1$), and the yaw ($x_6$) DOF. This is typical for a
  marine surface craft that is passively stable in roll and pitch. 
\end{itemize}
The three classes of plant differ only in their actuation.  We assume that all three plants share identical
mass and inertia  properties of $m_{11} = m_{22} = m_{33} = 11 \ kg$, $Ixx = 6.5 \ kg*m^2$, $Iyy = 5 \ kg*m^2$, and $Izz = 7 \ kg*m^2$.  
The control input, $\tau(t) \in {\rm I\!R}^6$ for Class 0 was 
\begin{align}
\tau_0(t) = \begin{vmatrix}
0.5*sin(t) \ N \\
0.5*sin(1.1*t) \ N \\
0.5*sin(0.9*t) \ N \\
cos(0.9*t) \ N \cdot m \\
sin(0.5*t) \ N \cdot m \\
cos(0.7*t) \ N \cdot m 
\end{vmatrix},
\label{eq.control-input-class-0}
\end{align}
and the control input for Class 1 was
\begin{align}
  \tau_1(t) = diag([1\; 0\; 0\; 1\; 1\; 1]) \tau_0(t),
  \label{eq.control-input-class-1}  
\end{align}
and the control input for Class 2 was
\begin{align}
  \tau_2(t) = diag([1\; 0\; 0\; 0\; 0\; 1]) \tau_0(t).
  \label{eq.control-input-class-2}
\end{align}

The adaptation gains of $\mathcal{A} = 100 \cdot I_{6 \times 6}$, and $\Gamma_1 = diag(6e3, 6e4, 6e4, 6e3, 6e3, 6e3)$ were selected empirically for reasonable performance.
The \ac{UCO} condition was evaluated with sliding time interval of $\delta = 20 \ sec$.   The discrete timestep was $.01 \ sec$.  The initial estimates of the inertial parameters were each initialized to be 1.4 times larger than the magnitude of the ``true'' values.
\subsection{Simulation Results}

These simulations are a preliminary  numerical investigation of the \acf{UCO} of mass and inertia parameters for the plant (\ref{simpleRBSystem}) and
the \ac{AID} defined in (\ref{v_update}) and (\ref{m_update}) for the cases of fully actuated and underactuated plants, Class 0, 1, and 2.
Each of the three trials used the control inputs specified in
(\ref{eq.control-input-class-0})
(\ref{eq.control-input-class-1})
and
(\ref{eq.control-input-class-2}), and identical initial estimates and adaptation gains.

\begin{figure}[h]
  \includegraphics[width=1\linewidth]{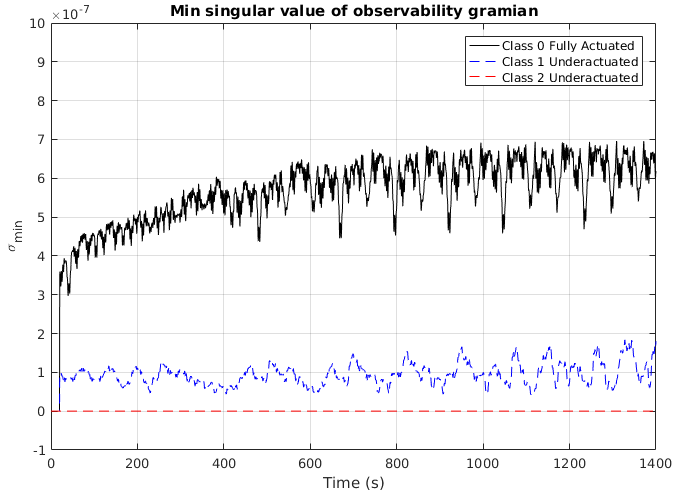}
  \caption{Plot of minimum singular values the observability Gramian (\ref{obsv_gram}) with $\delta = 20 \ sec$.}
  \label{fig:singvals}
\end{figure}

Figure \ref{fig:singvals} shows the evolution of the minimum singular value of the observability Gramian during three simulation trials of a rigid body model: Class 0 fully actuated, Class 1 underactuated, and Class 2 underactuated.  
As expected, the minimum singular value of the observability Gramian varied with the control authority available in the plant.
The minimum singular value of the observability Gramian was greatest for AID of the the Class 0 (fully actuated) plant. 
In the case of AID of the Class 1 underactuated plant, the minimum singular value of the observability Gramian was smaller in magnitude, but still clearly greater than zero.  It is clear that there exists some choice of control input, adaptation gain, and interval time $\delta$ to ensure the error dynamics of Class 1 plants are UCO. 
AID of the Class 2 underactuated plant produced minimum observability Gramian singular value that remained numerically zero during the entire simulation.
We can understand this intuitively, as Class 2 underactuated plants are constrained to planar motion, and thus some inertial parameters are not observable.   

The observable subspace is the image (span of the linearly-independent columns) of the observability matrix (\ref{eq:obsmatrix}).
The observable subspace exactly correlates with the results in Figure \ref{fig:singvals}.  For example, in the Class 2 plant the unobservable subspace corresponds to $\Delta m_{33}$, $\Delta m_{44}$, and $\Delta m_{55}$.  

\begin{figure}[h]
  \includegraphics[width=1\linewidth]{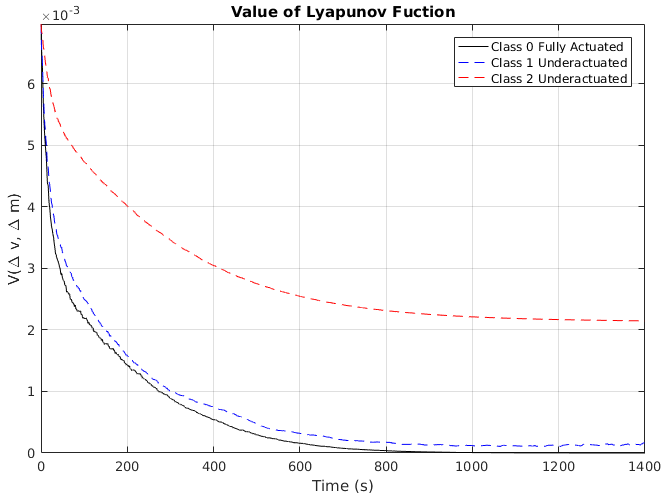}
  \caption{Plot of value of Lyapunov function during numerical simulation. }
  \label{fig:Lyapunov}
\end{figure}

The evolution of the value of the Lyapunov function also depended on control authority available in the degrees-of-freedom.  A comparison among the three simulations is shown in Figure \ref{fig:Lyapunov}.   As expected, the value of the Lyapunov function decreased fastest for the fully actuated plant.   At the end of the simulation, the value of the Lyapunov function for the Class 1 underactuated plant was approximately 1\% of the initial value.  This is partially a result of the fact that adaption gains were the same across all simulation studies in an effort to present a clear comparison between the performance of the AID in fully actuated and underactuated systems.  Clearly, in practice adaptation gains would be selected to best match the AID to a particular rigid body system.  In the case of underactuated rigid body systems - such as Class 1 systems - different adaptation gains should be used for better performance, including perhaps gain scheduling. 

\section{Conclusion}
\label{sec.conc}

This paper reported necessary and sufficient conditions for the \ac{UCO} of inertial parameters for second-order rigid-body dynamical systems given by (\ref{simpleRBSystem}), and the stable adaptive identifier for these class of plant models given by (\ref{v_update}) and (\ref{m_update}).  When the \ac{UCO} condition is satisfied, the adaptive parameter estimates are shown to converge to the true plant parameter values.
To the best of our knowledge this is the first reported analytical proof for this class of systems of UCO of plant parameters and for convergence of adaptive parameter estimates to true parameter values.  

Numerical simulations suggest that with the proper choice of control input, the UCO condition can be met for both fully-actuated plants and some classes of underactuated plants.

These results apply to a many 6-\ac{DOF} rigid-body dynamical plants; examples include aerospace, marine, and terrestrial robotic vehicles. We conjecture that this approach can be extended to include other parameters that appear in rigid body plant models including parameters for drag, buoyancy, added mass, bias, and actuators.  This is the focus of future work.


\balance
\bibliographystyle{sty/IEEEtranS.bst}
\bibliography{refs.bib,bib/llw.bib,bib/books.bib,bib/robot.bib,bib/rov.bib,bib/iros2011.bib}

\end{document}